\documentclass[twocolumn,superscriptaddress,showpacs]{revtex4}

\usepackage{amsmath}
\usepackage{amstext}
\usepackage{latexsym}
\usepackage{graphicx}

\newcommand{\ket}[1]{\left\vert#1\right\rangle}
\newcommand{\bra}[1]{\left\langle#1\right\vert}
\newcommand{\sprod}[2]{\left\langle#1\right\vert\left.#2\right\rangle}
\newcommand{\ad}{\hat{a}^\dag}
\newcommand{\bd}{\hat{b}^\dag}
\newcommand{\At}{\mathbf{\tilde{A}}}
\newcommand{\A}{\mathbf{A}}
\newcommand{\rr}{\mathbf{\b{r}}}
\newcommand{\M}{\mathbf{M}}
\newcommand{\Q}{\tilde{Q}}

\begin{document}

\title{On a linear optical implementation of non local product states and on their indistinguishability}
\author{Angelo Carollo}
\author{G.Massimo Palma}
\affiliation{Dipartimento di Scienze Fisiche ed Astronomiche and
INFM-Unit\`a di Palermo,\\ Via Archirafi 36, I-90123 Palermo,
Italy}
\author{Christoph Simon}
\affiliation{Centre for Quantum Computation, Clarendon Laboratory,
University of Oxford, Parks Road, Oxford OX1 3PU, United Kindom}
\author{Anton Zeilinger}
\affiliation{Institut f\"ur Experimentalphysik, Universit\"at
Wien, Boltzmanngasse 5, 1090 Wien, Austria}

\date{\today}
\begin{abstract}
 In a recent paper Bennett et al.~\cite{bennett} have shown the
 existence of a basis of product states of a bipartite system
 with manifest non-local properties. In particular these states cannot be
 completely discriminated by means of bilocal measurements. In this paper
 we propose an optical realization of these states and we will
 show that they cannot be completely discriminate by means of a
 global measurement using only optical linear elements,
 conditional transformation and auxiliary photons.
\end{abstract}

\maketitle

%%%%%%%%%%%%%%%%%%%%%%%%%%%%%%%%%%%%%%%%%%%%%%%%%%%%%%%%%%%%%%%%%%%%%%%%%%%%%
%\begin{multicols}
\section{Introduction}

Quantum optical systems are ideal  for the experimental test of
the foundation of quantum mechanics \cite{foundations} as well as
for  the experimental implementation of quantum information
protocols like quantum cryptography \cite{qcrypto}, quantum
teleportation \cite{teleport}, quantum dense coding~\cite{qdense}
and quantum computation~\cite{milburn}. In most of the above
experiments the key point is the generation and the detection of
entangled states. While the generation of various kind of
entangled states is now part of the daily routine of a good
laboratory the detection can be a surprisingly difficult task. The
most typical example is probably the detection of Bell states
\cite{bell}, for which it has been shown the impossibility to
build a setup able to discriminate with 100\% efficiency all the
four Bell states using only linear optical devices
\cite{vaidman,norbert,norbert2}. Such impossibility to
discriminate the states of an orthogonal basis is by no means
restricted to entangled systems. We will show that this difficulty
is present also in the case of an orthogonal basis of a bipartite
system which has been introduced in connection with non locality
without entanglement. Non locality has always been associated with
quantum entanglement. In a recent article however \cite{bennett}
Bennett et al have provided a counterexample by showing the
existence of an orthogonal set of states of a bipartite system
which,  although not entangled, are not distinguishable by means
of bilocal measurements. Given two particles, each of which
described by a three dimensional Hilbert space, they construct the
following orthogonal basis:

\begin{eqnarray}
\label{primi}
\ket{\psi_0}&=& \ket{2}_A\otimes\ket{2}\nonumber\\
\ket{\psi_{\pm 1}}&=&\frac{1}{\sqrt{2}}\ket{1}_A\otimes(\ket{1}\pm\ket{2})_B\nonumber\\
\ket{\psi_{\pm 2}}&=&\frac{1}{\sqrt{2}}\ket{3}_A\otimes(\ket{2}\pm\ket{3})_B\nonumber\\
\ket{\psi_{\pm 3}}&=&\frac{1}{\sqrt{2}}(\ket{2}\pm\ket{3})_A\otimes\ket{1}_B\nonumber\\
\ket{\psi_{\pm 4}}&=&\frac{1}{\sqrt{2}}(\ket{1}\pm\ket{2})_A\otimes\ket{3}_B
\end{eqnarray}

where A,B label the two particles and $\ket{1} \ket{2} \ket{3} $
are three orthogonal states for each particle.

The peculiar property of states (\ref{primi}) is that they cannot
be reliably distinguished by two separate observers by means of
any sequence of local operations even if they are allowed to
exchange classical communication.

In this paper we propose an optical realization of states
(\ref{primi})  and investigate the possibility to fully
discriminate them with a {\em global measurement by means of
linear elements}. A related problem has been investigated in
connection with the possibility to discriminate Bell states. It
has been shown \cite{vaidman,norbert,norbert2}that it is not
possible to perform a complete Bell measurement on a product
Hilbert space of two two-level bosonic systems states by means of
purely linear optical elements. One might expect that this is due
to the entangled nature of the Bell states. However, following the
line of~\cite{norbert}, we will show that also
states~(\ref{primi}) are not fully distinguishable by a global
measurement using only linear elements, even though they are not
entangled.
\begin{figure}[ht]\label{figdomino}
\centering
\includegraphics[width=6cm]{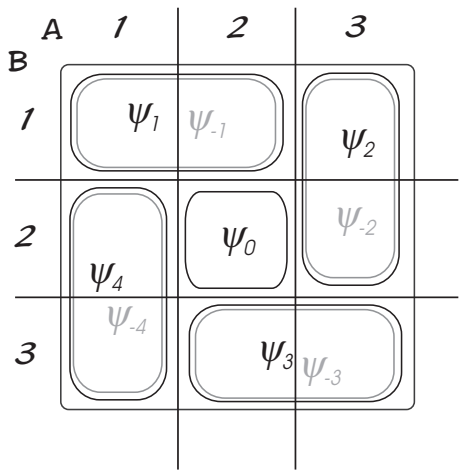}
\caption{Graphical representation of the states of
ref.~\cite{bennett} as a system of dominos. The fact that even if
these states are globally orthogonal their parts are not, is
evident in the picture, where the measurement are represented as a
cut along dashed lines.}
\end{figure}

\section{the setup}\label{setup}
In our optical setup the three dimensional Hilbert space of each
subsystem is mapped into the single photon state of three
different modes of the electromagnetic field. The  basis states
$\ket{1} \ket{2} \ket{3} $ for each of the two subsystems will
therefore be of the form $\ket{i}_A = a^{\dagger}_i \ket{0},
\ket{i}_B = b^{\dagger}_i \ket{0}$ where $a^{\dagger}_i ,
b^{\dagger}_i$ (1=1,2,3) are bosonic creation operators of three
orthogonal modes and $\ket{0}$ is the vacuum state. In this
notation states (\ref{primi}) are written as follows
 \begin{eqnarray}\label{stati}
  \ket{\psi_0}&=& \ad_2\bd_2\ket{0}\nonumber\\
  \ket{\psi_{\pm 1}}&=&\frac{1}{\sqrt{2}}\ad_1(\bd_1\pm\bd_2)\ket{0}\nonumber\\
  \ket{\psi_{\pm 2}}&=&\frac{1}{\sqrt{2}}\bd_1(\ad_3\pm\ad_2)\ket{0}\nonumber\\
  \ket{\psi_{\pm 3}}&=&\frac{1}{\sqrt{2}}\ad_3(\bd_3\pm\bd_2)\ket{0}\nonumber\\
  \ket{\psi_{\pm 4}}&=&\frac{1}{\sqrt{2}}\bd_3(\ad_1\pm\ad_2)\ket{0}
 \end{eqnarray}
The impossibility to distinguish states (\ref{stati})  by means of
bilocal measurements implies that they are not distinguishable by
measuring directly the photon number of each individual mode. A
first attempt to implement a collective measurement could be to
mix the modes by means of linear devices and than to measure the
output modes of such device. However, following \cite{norbert} we
will adopt  a more general strategy. We will assume to have at our
disposal a set of as many additional modes as we like,  here
indicated with bosonic creation operators $c^{\dagger}_j$, with
any number of photons we like and we will assume that these
auxiliary modes can be mixed with modes $a^{\dagger}_i,
b^{\dagger}_k$ in a black box.

\begin{figure}[ht]
\centering
\includegraphics[width=8cm]{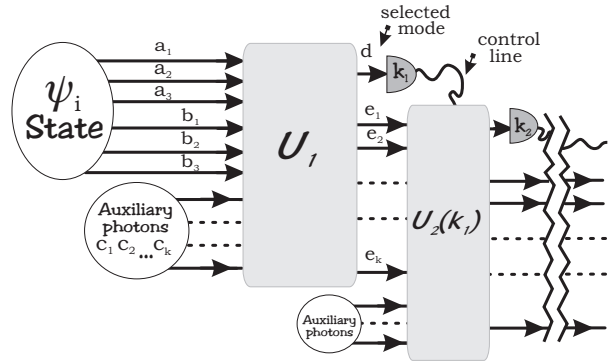}
\caption{Cascade setup in which the modes of the states
(\ref{stati}) are mixed in a first "box" with auxiliary modes.
Selected output mode is then measured and depending on its outcome
the remaining output modes are fed in a new box. The process can
be repeated over and over again}
\label{cascade}
\end{figure}

The output modes of this box are linked to the input ones by a
unitary transformation $U$. It has been shown \cite{zukowski} that
any such unitary transformations of modes can be obtained by means
of linear optical devices, like beam splitters and phase shifters.
To ensure the largest possible generality in our measurement
apparatus we will assume the possibility to perform conditional
measurements. In practice this means what follows: assume that a
measurement is made on one selected output mode while the other
are kept in a delay loop  and that, according to the outcome of
the measurement, these modes are fed into a selected further black
box, in a cascade setup ( see figure~\ref{cascade} ). The final
assumption we will make is that our detectors have the ability to
discriminate the number of incident photons. This assumption is
clearly unrealistic. We will show, however, that even if such
detectors were available, the measurement setup described above
cannot discriminate states~(\ref{stati}).

%\begin{widetext}
%\end{widetext}

%%%%%%%%%%%%%%%%%%%%%%%%%%%%%%%%%%%%%%%%%%%%%%%%%%%%%%%%%%%%%%%%%%%%%%%%%%%%%%
\section{Symmetry properties}
%%%%%%%%%%%%%%%%%%%%%%%%%%%%%%%%%%%%%%%%%%%%%%%%%%%%%%%%%%%%%%%%%%%

In this section we will describe some symmetry properties of
states (\ref{stati}) which are not only  interesting per se but
also will turn out useful in the following.

Consider the following transformation $  \mathbf{\hat{T}}$ which
permutes the modes of photon $A$ with the ones of photon $B$:
 \[
  \mathbf{\hat{T}}:\left\{\begin{array}{ccc}
    \ket{i}_A&\rightarrow&\ket{i}_B\\
    \ket{i}_B&\rightarrow&\ket{4-i}_A
   \end{array}\right.
 \]

This is obviously a linear transformation. In the basis states
$\ket{1}_A$,$\ket{2}_A$,$\ket{3}_A$,$\ket{1}_B$,$\ket{2}_B$,$\ket{3}_B$
$\mathbf{\hat{T}}$ takes the following matrix form

 \[
  \mathbf{\hat{T}}=\left(
  \begin{array}{cccccc}
   0&0&0&0&0&1\\
   0&0&0&0&1&0\\
   0&0&0&1&0&0\\
   1&0&0&0&0&0\\
   0&1&0&0&0&0\\
   0&0&1&0&0&0
  \end{array}\right)
 \]

The set of states~(\ref{stati}) is globally invariant under the
action of $\mathbf{\hat{T}}$ since

 \[
   \left\{
   \begin{array}{ccc}
    \ket{\psi_0}&\stackrel{\mathbf{\hat{T}}}{\longrightarrow}&\ket{\psi_0}\\
    \ket{\psi_{\pm 1}}&\stackrel{\mathbf{\hat{T}}}{\longrightarrow}&\ket{\psi_{\pm 2}}\\
    \ket{\psi_{\pm 2}}&\stackrel{\mathbf{\hat{T}}}{\longrightarrow}&\ket{\psi_{\pm 3}}\\
            \ket{\psi_{\pm 3}}&\stackrel{\mathbf{\hat{T}}}{\longrightarrow}&\ket{\psi_{\pm 4}}\\
    \ket{\psi_{\pm 4}}&\stackrel{\mathbf{\hat{T}}}{\longrightarrow}&\ket{\psi_{\pm 1}}\\
   \end{array}\right.
 \]

furthermore it is straightforward to verify that
$\mathbf{\hat{T}}^4 =\mathbf{\hat{1}}$. Another linear
transformation we will use in the following is the one which
introduces a phase change of $\pi$ on states $\ket{2}_A$ and
$\ket{2}_B$ leaving unaltered all the others. In matrix form

 \[
   \mathbf{\hat{S}}=\left(
   \begin{array}{rrrrrr}
    1&0&0&0&0&0\\
    0&-1&0&0&0&0\\
    0&0&1&0&0&0\\
    0&0&0&1&0&0\\
    0&0&0&0&-1&0\\
    0&0&0&0&0&1
   \end{array}\right)
 \]

The action of   $\mathbf{\hat{S}}$ on states~(\ref{stati}) is
simply

\[
  \bf{\hat{S}}:\ket{\psi_i}\rightarrow\ket{\psi_{- i}}
 \]

With   $ \mathbf{\hat{S}},\mathbf{\hat{T}}$ form a group which
leaves states~(\ref{stati}) invariant. Furthermore, by repeated
action of $ \mathbf{\hat{S}}$ and $\mathbf{\hat{T}}$, it is
possible to transform any $\ket{\psi_i}$ into any other
$\ket{\psi_j}$, with the exception of $\ket{\psi_0}$ which is
mapped onto itself. For instance we can transform $\psi_1$ into a
generic $\psi_{\pm k}$ (with k=1...4) by acting with the operator

 \[
  \mathbf{\hat{R}_{\pm k}}=\mathbf{\hat{S}}^{\frac{1\pm 1}{2}}\cdot\mathbf{\hat{T}}^{k-1}
 \]

This implies that the problem of how to generate the
states~(\ref{stati}) reduces to the problem of how to generate one
of them as the others can be obtained by repeated action of $
\mathbf{\hat{S}}$ and $\mathbf{\hat{T}}$ and, as we have said
already, this can be achieved by linear optical devices.

%%%%%%%%%%%%%%%%%%%%%%%%%%%%%%%%%%%%%%%%%%%%%%%%
\section{Auxiliary photons do not increase distinguishability}
%%%%%%%%%%%%%%%%%%%%%%%%%%%%%%%%%%%%%%%%%%%%%%%%
We will now show that the use of auxiliary photons in the
measurement setup described in section~\ref{setup} does not help
in increasing the distinguishability of states~(\ref{stati}). The
argument is a generalization to our more complex set of states of
the one used in~\cite{norbert} in connection with the problem of
distinguishing Bell states  with an analogous setup. In this
section we will outline the proof, leaving the details to
appendix~\ref{appA}.

As described already our measuring apparatus consists of a cascade
of "black boxes", in which modes are linearly mixed, and partial
measurements, which determine the sequence of unitary mixing. The
first of such black box, denoted by  $U_1$, is made out of linear
optical elements and its input and output are a set of bosonic
modes. The joint input modes consist of our six "system" modes
$a^{\dagger}_i, b^{\dagger}_k$ and an arbitrary number of
auxiliary modes $c^{\dagger}_i$. These input modes are unitarily
mixed in the box into a set of output modes $e^{\dagger}_i,
d^{\dagger}$ where the $d^{\dagger}$  mode is the one on which a
measurement will be performed. The measurement outcome will
determine the specific unitary mixing that will be performed in
next step of the measurement, consisting of a second box $U_2$.
While the measurement on mode $d^{\dagger}$ is performed the
photons in the remaining $e^{\dagger}_i$ modes are kept in a
waiting loop. The whole measurement procedure consists of a
cascade of conditional measurements as described above.

Let's now look more in detail at the first block of the apparatus.
The input state of $U_1$ can be written as

\begin{equation}
\ket{\psi_i^{tot}} = \ket{\psi_{aux}}\otimes\ket{\psi_i} =
P_{aux}(c^{\dagger}_k) P_i(a^{\dagger}_n, b^{\dagger}_m)\ket{0}
\label{input}
\end{equation}

where $P_i(a^{\dagger}_i, b^{\dagger}_i)$ is a polynomial of
degree 2 and $P_{aux}(c^{\dagger}_k)$ is a polynomial of arbitrary
degree in the $c^{\dagger}_k$

The corresponding
output state is

\begin{equation}
\ket{\psi^{tot}_i}=\tilde{P}_{aux}(d^\dag,e^\dag_k)\tilde{P}_{\psi_i}(d^\dag,e^\dag_k)\ket{0}
\label{output}
\end{equation}

Where $\tilde{P}_{aux}(d^\dag,e^\dag_k)$ and
$\tilde{P}_{\psi_i}(d^\dag,e^\dag_k)$ are nothing but
$P_{aux}(c^{\dagger}_k)$ e $P_i(a^{\dagger}_n,
b^{\dagger}_m)\ket{0}$ written in terms of the creation and
annihilation operators at the output of $U_1$.

We can expand ${\tilde P}_{aux}, {\tilde P}_{\Psi_i}$ in terms of
decreasing powers of $d^{\dagger}$ as follows

\begin{align}
\tilde{P}_{aux}(d^\dag,e^\dag_k)&=
\sum_{n=0}^{n_a}(d^\dag)^n\Q_a^{(n)}(e^\dag_k)\\
\tilde{P}_{\psi_i}(d^\dag,e^\dag_k)&=
\sum_{n=0}^{n_s}(d^\dag)^n\Q_{\psi_i}^{(n)}(e^\dag_k)\label{espan}
\end{align}

In (\ref{espan}) $n_s$ is the largest order in $d^{\dagger}$ for
the nine ${\tilde P}_{\Psi_i}$ and by definition is independent on
index $i$ (${\tilde Q}_{\Psi_i}$ can be zero for some $i$).
Analogously $n_a$ is defined as the order in $d^{\dagger }$ of
polynomial  ${\tilde P}_{aux}$. We can therefore rewrite
(\ref{output}) as

\begin{equation}
\ket{\psi^{tot}_i}=\sum_{n,m=0}^{n_a,n_s}(d^\dag)^{n+m}\Q_a^{(n)}(e^\dag_n)\Q_{\psi_i}^{(m)}(e^\dag_k)\ket{0}
\end{equation}

Out of the possible outcomes of the measurement of the number $N$
of photons in mode $d$ we will concentrate on two particular
outcomes, namely those resulting in the highest number, $N_{max}$
and $N_{max}-1$ where $N_{max}= n_s + n_a$. The reason of this
particular choice will be shortly evident.

Let's suppose now that the number of photons on the selected mode
$d$ is measured. If $N$ is the outcome of such measurement  the
(unormalised) conditional state of the remaining modes can be we
written as

\begin{equation}\label{stcond}
\ket{\psi^{cond\rightarrow N}_i}=\sum_{n=max\{0,N-n_s\}}^{min\{n_a,N\}}\Q_a^{(n)}\Q_{\psi_i}^{(N-n)}\ket{0}
\end{equation}

If the input states are to be distinguishable the conditional
states $\ket{\psi^{N}_i}$ \emph{must be orthogonal for each
possible value of $N$}, i.e.
\[
\sprod{\psi_i^{N}}{\psi_j^{N}} = 0 \hspace{0.5cm} \forall N, i\neq j
\]

In appendix~\ref{appA} we will show that the two conditions

\begin{equation}\label{withaux}
 \left\{
  \begin{array}{l}
   \sprod{\psi_i^{N_{max}}}{\psi_j^{N_{max}}}=0\\
   \sprod{\psi_i^{N_{max}-1}}{\psi_j^{N_{max}-1}}=0
  \end{array}
 \right.
\end{equation}

can be simultaneously satisfied if and only if the two conditions

\begin{equation} \label{without}
 \begin{cases}
   \bra{0} \Q_{\psi_i}^{(n_s)\dag}\Q_{\psi_j}^{(n_s)}\ket{0}=0\\
   \bra{0} \Q_{\psi_i}^{(n_s-1)\dag}\Q_{\psi_j}^{(n_s-1)}\ket{0}=0\quad(\mbox{for } n_s\ne 0)
 \end{cases}
\end{equation}

are simultaneously satisfied. The important point is
that~(\ref{without}) {\em do not depend on the auxiliary input
states}. It is easy to convince oneself that this is the case
since from~(\ref{stcond})  follows that

\begin{equation}
\bra{0} \Q_{\psi_i}^{(N)\dag}\Q_{\psi_j}^{(N)}\ket{0}
\propto \sprod{\psi_i^{N}}{\psi_j^{N}}_{n_a=0}
\end{equation}

where $\ket{\psi_i^N}_{n_a=0}$ is the conditional output state
obtained from $\psi_i$ in the absence of auxiliary photons when
$N$ photons are measured in mode  $d$.

The central point of this section is that the fact that
condition~(\ref{withaux}) implies condition~(\ref{without}) is
equivalent to say that any pair of states $\psi_i$, $\psi_j$ are
distinguishable in the presence of auxiliary photons only if they
are distinguishable in the absence of auxiliary photons. In other
words auxiliary photons do not improve complete
distinguishability.

%%%%%%%%%%%%%%%%%%%%%%%%%%%%%%%%%%%%%
\section{It is impossible to build a  complete linear discriminator}
%%%%%%%%%%%%%%%%%%%%%%%%%%%%%%%%%%%%%
We will now show that it is impossible for states~(\ref{stati}) to
satisfy

\begin{subequations}
\begin{align}
   \sprod{\psi_i^{n_s}}{\psi_j^{n_s}}_{n_a=0}&=0 \label{condo1}\\
  \sprod{\psi_i^{n_s-1}}{\psi_j^{n_s-1}}_{n_a=0}&=0 \quad(\mbox{for } n_s \ne 0) \label{condo2}
\end{align}
\end{subequations}

for all $i$, $j \in \{-4..4\}$ ($i\ne j$ ).

In the absence of auxiliary photons states $\ket{\psi_i}$ can be
written in terms of a polynomial of creation operators as

\[
\ket{\psi_i}=P_{\psi_i}\left(a^\dag_1,a^\dag_2,a^\dag_3,b^\dag_1,b^\dag_2,b^\dag_3\right)\ket{0}
\]

Let's now define the creation operator vector as

\[
\A\equiv
\left(
\ad_1,\ad_2,\ad_3,\bd_1,\bd_2,\bd_3,\{c^\dag_k\}
\right)^T
\]

where the  $\{c^\dag_k\}$ are a possible set of (empty) auxiliary
modes. Since the  $\psi_i$ are two photon states they can be
written in terms of a real symmetric matrix $\M^{(i)}$ as follows
\[
\ket{\psi_i}=\A^T\M^{(i)}\A\ket{0}
\]

where the exact form of $\M^{(i)}$ can be obtained from~(\ref{stati}).

If $\mathbf{U}$ is a generic unitary matrix transforming the input modes into the output ones of our
apparatus than

\begin{equation}\label{psi}
\ket{\psi_i}=\At^T\mathbf{\tilde{M}}^{(i)}\At\ket{0}
\end{equation}

where
\[
\tilde{\M}^{(i)}=\mathbf{U}^T\M^{(i)}\mathbf{U}
\]
and
\[
\At=\mathbf{U}^\dag\A = \left( d^\dag,e^\dag_1,e^\dag_2,\dots
\right)^T
\]
with $d^\dag$ corresponding to the detected output mode.

States~(\ref{psi}) can than be written as

\begin{multline}\label{svlp}
\ket{\psi_i}=\tilde{M}_{00}^{(i)}(d^\dag)^2\ket{0}+\\
+2\sum_{k=1}^D\tilde{M}_{0k}^{(i)}d^\dag e^\dag_k\ket{0}+
\sum_{k,l=1}^D\tilde{M}_{kl}^{(i)}e^\dag_k e^\dag_l\ket{0}
\end{multline}

where $\tilde{M}^{(i)}_{kl}$ is the generic matrix element of $\tilde{\M}^{(i)}$ whose dimension $D+1$
corresponds to the number of output modes involved.

Let's write $\mathbf{U}$ as
\[
\mathbf{U}=\left(
\begin{array}{cc}
u_0&\mathbf{\b{r}}_0\\
u_1&\mathbf{\b{r}}_1\\
\vdots&\vdots\\
u_D&\mathbf{\b{r}}_D
\end{array}
\right)
\]

where  $u_i$ are the element of the first column of the matrix and
$\mathbf{\b{r}}_i$ ( with $i\in\{0..D\}$) are vectors of dimension
$D$ representing the remaining elements of row $i^{th}$. As a
consequence of the unitarity of $\mathbf{U}$ we have
\begin{equation}\label{uni}
u_i^*u_j+\rr_i^\dag\cdot\rr_j=\delta_{ij}
\end{equation}
We define the columns vector $\mathbf{\b{c}_0}$ whose elements are
the first 6 elements of the $0^{th}$ column of $\mathbf{U}$:
\begin{equation}\label{col}
\mathbf{\b{c}_0}=\left(
u_0,\dots,u_5,0,\dots,0
\right)^T
\end{equation}
We recall that $n_s$ is the highest degree of  $d^\dag$ in
polynomials $\At^T\tilde{\M^{(i)}}\At$ for all values of $i$, in
other words the maximum number of photons which can be detected in
$d$ for all possible input states $\{\psi_i\text{, }i\in\{-4\dots
4\} \}$. Obviously   $n_s$ can assume only values $0,1,2$,
depending on the specific choice of $\mathbf{U}$ and $d$. We will
now show that for all possible value of $n_s$ it is impossible to
satisfy simultaneously (\ref{condo1}) and~(\ref{condo2}).

\emph{$n_s=0$}: this corresponds to a bad choice of mode $d$, as
the monitored mode would be decoupled  from the input ones for all
possible input state.

\emph{$n_s=1$}: this corresponds to $\tilde{M}_{00}^{(i)}=0$ for
all value of $i$ (see~(\ref{svlp})). This implies that
\begin{equation}\label{ns1}
\begin{split}
\tilde{M}_{00}^{(i)}&=\sum_{k,l=0}^DM_{kl}^{(i)}u_k^*u_l=\\
&=\mathbf{\b{c}_0}^T\cdot\M^{(i)}\cdot\mathbf{\b{c}_0}=0
\end{split}
\qquad\forall\;i\in\{-4..4\}
\end{equation}

The above relation is a constrain on $\mathbf{\b{c}_0}$ which we
will now show to be incompatible with~(\ref{condo1}).

To this end we note that from~(\ref{svlp}) follows that after the
detection of one photon in mode $d$ the remaining modes are left
in the (unormalised) conditional state
\begin{align}\label{statocondns1}
\ket{\psi^{cond \rightarrow 1}_i} &=\sum_{k=1}^D\tilde{M}^{(i)}_{0k}e^\dag_k\ket{0}\nonumber\\
&=\boldsymbol{\vec{M}}_0^{(i)}\cdot\At\ket{0}
\end{align}

where for convenience of notation we have introduced the vector
\begin{equation}\label{vettM0}
\boldsymbol{\vec{M}}_0^{(i)}=
\sum_{k,l=1}^5 M_{kl}^{(i)}u_k \rr_l
= (0,\tilde{M}_{01}^{(i)},\tilde{M}_{02}^{(i)},\dots,\tilde{M}_{0D}^{(i)})
\end{equation}

From~(\ref{statocondns1}) and~(\ref{vettM0}) follows that the
trivial solution  $\mathbf{\b{c}_0}=\mathbf{\b{0}}$ implies
$\ket{\psi^{cond \rightarrow 1}_i}=0$  $\forall i$, i.e. $n_s=0$.
We must therefore look for possible nontrivial solutions of
eq.~(\ref{ns1}) compatible with~(\ref{condo1}), which in this
particular case reads
\begin{equation}\label{condo1ns1}
\sprod{\psi_i^{cond \rightarrow 1}}{\psi_j^{cond \rightarrow 1}}
={\boldsymbol{\vec{M}}_0^{(i)}}^\dag\cdot\boldsymbol{\vec{M}}_0^{(j)}=0
\end{equation}

However, as shown in appendix \ref{appB} conditions~(\ref{ns1})
and~(\ref{condo1ns1})  are compatible only with the trivial
solution. This implies that it is not possible to build a complete
discriminator for $n_s=1$.

\emph{$n_s=2$} this corresponds a non zero probability to measure
two photons in mode $d$ for some $\psi_i$ which implies
$\tilde{M}_{00}^{(i)}\neq 0$ for at least one value of $i$. On the
other hand condition~(\ref{condo1}) can be satisfied in this
specific case if and only if $\tilde{M}_{00}^{(i)}\neq 0$  for at
most one value of $i$, which we will denote by  $i_o$.
Condition~(\ref{condo1}) than becomes
\begin{equation}\label{ns2}
\tilde{M}_{00}^{(i)}=\mathbf{\b{c}_0}^T\cdot\M^{(i)}\cdot\mathbf{\b{c}_0}=0
\qquad i\ne i_o
\end{equation}
and~(\ref{condo2}) becomes equivalent to
condition~(\ref{condo1ns1}). In order to complete our proof it
will therefore be enough to show that whatever the value of $i_o$
conditions~(\ref{ns2}) and~(\ref{condo1ns1}) cannot be
simultaneously satisfied. Suppose in particular that they are not
satisfied for  $i_o =1$, the symmetry analysis carried out in the
previous section immediately lead to the conclusion that they
cannot be satisfied by any other value $i_o$ (apart from  $i_o
=0$). We have shown that it is always possible to build a linear
operator $\mathbf{\hat{R}_i}$ which transforms $\psi_1$ into
$\psi_i$ ($i\neq 0$) and leaves the set of states $\{\psi_i\}$
globally invariant. If there were a linear operator $\mathbf{U'}$
such to satisfy conditions~(\ref{ns2}) and~(\ref{condo1ns1}) for
any value of $i_o \neq 0$ than
$\mathbf{U}=\mathbf{\hat{R}_i}^\star\cdot\mathbf{U'}\cdot\mathbf{\hat{R}_i}$
would satisfy the same conditions for $i_o =1 $, which contradicts
our initial assumption. The problem than reduces to the analysis
of the cases  $i_o =0$  and $i_o =1 $.  Such analysis,
straightforward but tedious (see appendix~\ref{appB}), leads to
the result that indeed for both values of $i_o$
conditions~(\ref{ns2}) and~(\ref{condo1ns1}) are incompatible.
%%%%%%%%%%%%%%%%%%%%%%%%%%%%%%%%
\section{conclusions}
%%%%%%%%%%%%%%%%%%%%%%%%%%%%%%%%%
In this paper we have proposed an optical realization of
states~(\ref{primi}). Bennett et al.~\cite{bennett} have shown
that they cannot be discriminated by means of local action and
classical communication. We have demonstrated that to add the
possibility of global interference it is still not sufficient. In
other words we have shown the impossibility to fully discriminate
them by means of a global measurement using linear elements, like
beam splitters and phase shifters, delay lines and electronically
switched linear elements, photodetectors and auxiliary photons.

The impossibility to implement such a measurement has already been
shown for the set of maximally entangled Bell states. We have
proved an analogous no-go theorem for a set of states which,
although non local, are not entangled. This opens new questions on
which class of photon states can be in general be fully
discriminated by means of linear optical systems.

%%%%%%%%%%%%%%%%%%
%%%%%%%%%%%%%%%%%%

\section*{Acknowledgments}
We would like to thank C.H. Bennett, J.Calsamiglia, D.DiVincenzo,
A.K.Ekert, I.Jex, N.L\"utkenhaus for helpful discussions. This
work was supported in part by the EU under grants TMR - ERB FMR
XCT 96-0087 - "The Physics of Quantum Information" IST - 1999 -
11053 - EQUIP,"Entanglement in Quantum Information Processing and
Communication".

%%%%%%%%%%%%%%%%%%%
%%%%%%%%%%%%%%%%%%%

\appendix

%%%%%%%%%%%%%%%%%%%
%%%%%%%%%%%%%%%%%%%

\section{}
\label{appA} In this appendix we will show that the two conditions
\begin{subequations}\label{equivconditions}
\begin{equation}\label{equivconditions1}
 \left\{
  \begin{array}{l}
   \sprod{\psi_i^{N_{max}}}{\psi_j^{N_{max}}}=0\\
   \sprod{\psi_i^{N_{max}-1}}{\psi_j^{N_{max}-1}}=0
  \end{array}
 \right.
\end{equation}
can be simultaneously satisfied if and only if the following
conditions
\begin{equation}\label{equivconditions2}
 \left\{
  \begin{array}{l}
   \bra{0} \Q_{\psi_i}^{(n_s)\dag}\Q_{\psi_j}^{(n_s)}\ket{0}=0\\
   \bra{0} \Q_{\psi_i}^{(n_s-1)\dag}\Q_{\psi_j}^{(n_s-1)}\ket{0}=0
  \end{array}
 \right.
\end{equation}
\end{subequations}
are simultaneously satisfied. From~(\ref{stcond}) follows that the
scalar product between the (unormalised) states
$\ket{\psi^{cond\rightarrow N}_i}, \ket{\psi^{cond\rightarrow
N}_j}$ obtained after the measurement of $N$ photons in mode $d$
is
\begin{equation}\label{prodN}
\sprod{\psi^N_i}{\psi^N_j}=
\sum_{n,m}
\bra{0}\Q_{\psi_i}^{(N-m)\dag}
\Q_a^{(m)\dag}
\Q_a^{(n)}
\Q_{\psi_i}^{(N-n)}\ket{0}
\end{equation}

with $max\{0,N-n_s\}\leq n,m \leq min\{n_a,N\}$.

Let's first consider the case $N=n_a+n_s=N_{max}$ :

\begin{equation}
\label{prodNmax}
 \begin{split}
 &\quad\sprod{\psi_i^{N_{max}}}{\psi_j^{N_{max}}}=\\
 &=\bra{0}\Q_{\psi_i}^{(n_s)\dag}
  \Q_a^{(n_a)\dag}
  \Q_a^{(n_a)}
  \Q_{\psi_j}^{(n_s)}\ket{0}\\
 &=\bra{0}\Q_a^{(n_a)\dag}
  \Q_a^{(n_a)}
  \Q_{\psi_i}^{(n_s)\dag}
  \Q_{\psi_j}^{(n_s)}\ket{0}\\
 &=\sum_{\{\mathbf{n}\}}
  \bra{0}\Q_a^{(n_a)\dag}
  \Q_a^{(n_a)}
  \ket{\mathbf{n}}\bra{\mathbf{n}}
  \Q_{\psi_i}^{(n_s)\dag}
  \Q_{\psi_j}^{(n_s)}
  \ket{0}\\
 &=\bra{0}\Q_a^{(n_a)\dag}
  \Q_a^{(n_a)}
  \ket{0}\bra{0}\Q_{\psi_i}^{(n_s)\dag}
  \Q_{\psi_j}^{(n_s)}
  \ket{0}
 \end{split}
\end{equation}
Above we have used the fact that
$\left[\Q_{\psi_i}^{(n_s)},\Q_a^{(n_a)\dag}\right]\!\!=\!0$~\cite{commut}
and introduced the completeness relation
$\sum_{\{\mathbf{n}\}}\ket{\mathbf{n}}\bra{\mathbf{n}}$, where
$\ket{\mathbf{n}}$ is a Fock states of the relevant modes. Note
that only the term corresponding to $\ket{0}\bra{0}$ survives.

Let's now evaluate~(\ref{prodN}) when $N=N_{max}-1$
\begin{equation}\label{somma}
\sprod{\psi_i^{N_{max}-1}}{\psi_j^{N_{max}-1}}
=\sum_{n,m=0}^1{\cal C}_{m,n}(i,j)
\end{equation}

where
\begin{multline*}
{\cal C}_{m,n}(i,j)=\\
=\bra{0}\Q_{\psi_i}^{(n_s-m)\dag}
\Q_a^{(n_a-1+m)\dag}
\Q_a^{(n_a-1+n)}
\Q_{\psi_j}^{(n_s-n)}\ket{0}
\end{multline*}

It is straightforward to verify \cite{commut} that
$\left[\tilde{P}_{aux},\tilde{P}_{\psi_i}^\dag\right]=0$ implies
that
\begin{subequations}\label{regoledicomm}
 \begin{equation}\label{comm}
  \left[\Q_a^{(n_a)},\Q_{\psi_i}^{(n)\dag}\right]=\left[\Q_a^{(m)},\Q_{\psi_i}^{(n_s)\dag}\right]=0\quad\forall\;m,n
 \end{equation}
and that:
 \begin{equation}
  \left[\Q_a^{(n_a-1)},\Q_{\psi_i}^{(n_s-1)\dag}\right]=
  n_a n_s\cdot\Q_a^{(n_a)}\Q_{\psi_i}^{(n_s)\dag}\label{comm1}
 \end{equation}
\end{subequations}

Relation~(\ref{comm}), with a procedure analogous to the one used
to derive~(\ref{prodNmax}), can be used to show that
\begin{align}
\label{c00}
 {\cal C}_{0,0}(i,j)&= \bra{0}\Q_{\psi_i}^{(n_s)\dag}
 \Q_a^{(n_a-1)\dag} \Q_a^{(n_a-1)} \Q_{\psi_j}^{(n_s)}\ket{0}\\
& =\bra{0}\Q_a^{(n_a-1)\dag} \Q_a^{(n_a-1)}\ket{0}\bra{0}
 \Q_{\psi_i}^{(n_s)\dag} \Q_{\psi_j}^{(n_s)}\ket{0}\nonumber
\end{align}

Let's now consider terms
\begin{eqnarray}
 {\cal C}_{1,0}(i,j) &=&
 \bra{0}\Q_{\psi_i}^{(n_s-1)\dag}\Q_a^{(n_a)\dag}
 \Q_a^{(n_a-1)} \Q_{\psi_j}^{(n_s)}\ket{0}\nonumber\\
 &=&\left(\bra{0}\Q_{\psi_j}^{(n_s)\dag} \Q_a^{(n_a-1)\dag}
 \Q_a^{(n_a)}  \Q_{\psi_i}^{(n_s-1)}\ket{0}\right)^*\nonumber\\
& =& {\cal C}^*_{0,1}(j,i)
\end{eqnarray}

which, with the help of~(\ref{comm}) can be expressed as
\begin{eqnarray}
\label{c10comm}
{\cal C}_{1,0}(i,j) &=& \bra{0} \Q_a^{(n_a)\dag}
  \Q_a^{(n_a-1)} \Q_{\psi_i}^{(n_s-1)\dag}
  \Q_{\psi_j}^{(n_s)} \ket{0}\\
& -&\bra{0} \Q_a^{(n_a)\dag} \left[\Q_a^{(n_a-1)},\Q_{\psi_i}^{(n_s-1)\dag}\right]
  \Q_{\psi_j}^{(n_s)} \ket{0}\nonumber
\end{eqnarray}

As all the states $\psi_i$ contain a definite number of photons,
namely  ${\cal N}=2$, $\tilde{P}_{\psi_i}(d^\dag,e_k^\dag)$ is a
homogeneous  polynomial of degree ${\cal N}$ in $d^\dag$ and
$e_k^\dag$ and therefore the generic  $\Q_{\psi_i}^{(n)}$ is a
homogeneous  polynomial of degree ${\cal N}-n$ in $e_k^\dag$. As a
consequence
$\Q_{\psi_i}^{(n_s-1)\dag}\Q_{\psi_j}^{(n_s)}\ket{0}=0$. From this
follows that the first term at the right hand side
of~(\ref{c10comm}) is equal to zero.

Finally, with the help of~(\ref{comm1}) and~(\ref{comm}) we obtain
\begin{align}\label{c10}
 {\cal C}_{1,0}(i,j)&=-n_a n_s
 \bra{0}\Q_a^{(n_a)\dag}
 \Q_a^{(n_a)}
 \Q_{\psi_i}^{(n_s)\dag}
 \Q_{\psi_j}^{(n_s)}\ket{0}\nonumber
\\
 &=-n_a n_s\bra{0}\Q_a^{(n_a)\dag}
 \Q_a^{(n_a)}
 \ket{0}\bra{0}
 \Q_{\psi_i}^{(n_s)\dag}
 \Q_{\psi_j}^{(n_s)}\ket{0}\nonumber
\\
 &={\cal C}^*_{0,1}(j,i)={\cal C}_{0,1}(i,j)
\end{align}

where again we have made use of a completeness relation.

We are left with the term ${\cal C}_{1,1}(i,j)$ in the sum of
eq.(\ref{somma}), which can be simplified with the same procedure
as in eq.(\ref{prodNmax}) to obtain
\begin{eqnarray}
\label{c11}
{\cal C}_{1,1}(i,j) &=& \bra{0} \Q_{\psi_i}^{(n_s-1)\dag}
 \Q_a^{(n_a)\dag} \Q_a^{(n_a)} \Q_{\psi_j}^{(n_s-1)}\ket{0}\\
& = &\bra{0}\Q_a^{(n_a)\dag} \Q_a^{(n_a)}\ket{0}\bra{0}
 \Q_{\psi_i}^{(n_s-1)\dag} \Q_{\psi_j}^{(n_s-1)}\ket{0}\nonumber
\end{eqnarray}

By inserting~(\ref{c00},~\ref{c10},~\ref{c11}) into~(\ref{somma})
we obtain
\begin{eqnarray}
\label{prodNmax-1}
 \sprod{\psi_i^{N_{max}-1}}{\psi_j^{N_{max}-1}}&= &
 {\cal A}_{n_s}\bra{0} \Q_{\psi_i}^{(n_s)\dag}
 \Q_{\psi_j}^{(n_s)}\ket{0}\\
 &+&{\cal A}_{n_s-1}\bra{0} \Q_{\psi_i}^{(n_s-1)\dag} \Q_{\psi_j}^{(n_s-1)}\ket{0}\nonumber
\end{eqnarray}

where, up to irrelevant multiplicative constants,
\begin{eqnarray}
{\cal A}_{n_s}&=&
 \bra{0}\Q_a^{(n_a-1)\dag}\Q_a^{(n_a-1)}\ket{0}-\nonumber\\
&-&2n_a n_s\bra{0}\Q_a^{(n_a)\dag}\Q_a^{(n_a)}\ket{0}\nonumber\\
 {\cal A}_{n_s-1} &=& \bra{0}\Q_a^{(n_a)\dag}\Q_a^{(n_a)}\ket{0}
\end{eqnarray}

which concludes our proof, as, from~(\ref{prodNmax})
and~(\ref{prodNmax-1}) follows the implication between
~(\ref{equivconditions1}) and~(\ref{equivconditions2}).

%%%%%%%%%%%%%%%%%%%%%%%%%%%%%%%
\section{}
\label{appB}
%%%%%%%%%%%%%%%%%%%%%%%%%%%%%%%%
In this appendix we will prove that condition~(\ref{ns2})  is
incompatible with~(\ref{condo1ns1}) for both $i_0=0$ and $i_0=1$.
To this goal it is helpful to define a matrix $\M$, linear
combination of the $\M^{(i)}$:
\[
\M=\sum_i\mu_i\M^{(i)}
\]

so that a generic input state can be defined as
\[
\ket{\psi}=\sum_i\mu_i\ket{\psi_i}=\A^T\mathbf{M}\A\ket{0}
\]

From~(\ref{stati})  follows that
\begin{widetext}
\[
 \mathbf{M}=\frac{1}{\sqrt{2^3}}\left(
 \begin{array}{*8{c}}
0&0&0&\mu_{-1}+\mu_1&\mu_{-1}-\mu_1&\mu_{-4}+\mu_4&\cdots&0\\
0&0&0&\mu_{-3}+\mu_3&\sqrt{2}\mu_1&\mu_{-4}-\mu_4&\cdots&0\\
0&0&0&\mu_{-3}-\mu_3&\mu_{-2}+\mu_2&\mu_{-2}-\mu_2&\cdots&0\\
\mu_{-1}-\mu_1&\mu_{-3}+\mu_3&\mu_{-3}-\mu_3&0&0&0&\cdots&0\\
\mu_{-1}-\mu_1&\sqrt{2}\mu_1&\mu_{-2}+\mu_2&0&0&0&\cdots&0\\
\mu_{-4}+\mu_4&\mu_{-4}-\mu_4&\mu_{-2}-\mu_2&0&0&0&\cdots&0\\
\vdots&\vdots&\vdots&\vdots&\vdots&\vdots&\ddots&\vdots\\
0&0&0&0&0&0&\cdots&0
 \end{array}
 \right)
\]
\end{widetext}

We can then write
\begin{eqnarray}
\label{m00}
 \tilde{M}_{00}&=&\mathbf{\b{c}_0^T}\cdot\mathbf{M}\cdot\mathbf{\b{c}_0}\nonumber\\
 &=&\mu_0\cdot u_1u_4+\\
 &+&\{\mu_1\cdot u_0(u_3+u_4)+\mu_{-1}\cdot u_0(u_3-u_4)+\nonumber\\
 &+&\mu_2\cdot u_2(u_4+u_5)/+\mu_{-2}\cdot u_2(u_4-u_5)+\nonumber\\
 &+&\mu_3\cdot u_3(u_1+u_2)/+\mu_{-3}\cdot u_3(u_1-u_2)+\nonumber\\
 &+&\mu_4\cdot u_5(u_0+u_1)/+\mu_{-3}\cdot u_5(u_0-u_1)\}/\sqrt{2}\nonumber
\end{eqnarray}

We now impose condition~(\ref{ns2}) with $i_0=0$ on vector
$\mathbf{\b{c}_0}$, i.e. we impose that $\tilde{M}_{00}^{(0)}$ is
the only nonzero coefficient. We have therefore to equal to zero
all coefficients in~(\ref{m00})  except the one multiplying
$\mu_0$. The only solution compatible with this condition is
\begin{equation}
\label{c0withI0}
\mathbf{\b{c}_0}=(0,u_1,0,0,u_4,0,\dots,0)^T
\end{equation}

From the form of $\M$ and from~(\ref{c0withI0}) follows that~(\ref{vettM0})
can  be rewritten as follows:
\begin{eqnarray}
\boldsymbol{\vec{M}}_0 &=&\sum_i\mu_i\boldsymbol{\vec{M}}_0^{(i)}\nonumber\\
&=&  \mu_0\cdot \frac{u_1\rr_4+\rr_1u_4}{2}
+\frac{\mu_1-\mu_{-1}}{2^{\frac{3}{2}}}u_4\rr_0+\\
&+&\frac{\mu_2+\mu_{-2}}{2^{\frac{3}{2}}}u_4\rr_2+\frac{\mu_3+\mu_{-3}}{2^{\frac{3}{2}}}u_1\rr_3
  +\frac{\mu_4-\mu_{-4}}{2^{\frac{3}{2}}}u_1\rr_5\nonumber
\end{eqnarray}

Condition~(\ref{condo1ns1}) implies
\begin{subequations}
\label{ns2x1}
\begin{align}
\sprod{\psi_1^{cond\rightarrow 1}}{\psi_{-1}^{cond\rightarrow 1}}\propto|u_4|^2\|\rr_0\|^2=0\\
\sprod{\psi_3^{cond\rightarrow 1}}{\psi_{-3}^{cond\rightarrow
1}}\propto|u_1|^2\|\rr_3\|^2=0
\end{align}
\end{subequations}

Since condition~(\ref{uni}) requires that
$\|\rr_0\|^2=\|\rr_3\|^2=1$, to fulfill ~(\ref{ns2x1}) we must
impose $|u_1|=|u_4|=0$. This however would imply $n_s=0$.

We will now show that conditions~(\ref{condo1ns1}),~(\ref{ns2})
cannot be simultaneously fulfilled with $i_0=1$, i.e that the only
non zero coefficient of $\tilde{M}_{00}$ is the one multiplying
the coefficient $\mu_1$.

Along the same lines of the previous case we obtain a constrain on
the vector $\mathbf{\b{c}_0}$ leading to the following relation
\[
\mathbf{\b{c}_0}=(u_0,0,0,u,u,0,\dots,0)^T
\]

and $\boldsymbol{\vec{M}}_0$ reduces to
\begin{eqnarray}
\boldsymbol{\vec{M}}_0
 &=&2^{-\frac{2}{3}}\{\mu_1 [u_0(\rr_3+\rr_4)+\rr_12u]+\nonumber\\
  &+&\mu_{-1} u_0(\rr_3-\rr_4) +(\mu_2-\mu_{-2}) u\rr_2+\nonumber\\
&  +&(\mu_4+\mu_{-4}) u_0\rr_5  +\mu_3 u(\rr_1+\rr_2)+\\
 & +&\mu_{-3} u(\rr_1-\rr_2)\}\nonumber
\end{eqnarray}

Therefore we obtain
\begin{align*}
&\sprod{\psi_2^{cond \rightarrow 1}}{\psi_{-2}^{cond \rightarrow 1}}\propto|u|^2\|\rr_2\|^2=0\\
&\sprod{\psi_4^{cond \rightarrow 1}}{\psi_{-4}^{cond \rightarrow
1}}\propto|u_0|^2\|\rr_5\|^2=0
\end{align*}

From the unitarity condition~(\ref{uni}) follows that $\|\rr_2\|=\|\rr_5\|=1$ and~(\ref{condo1}) can be satisfied
only if $u$ and $u_0$  are both zero.

As before this requirement leads to the trivial solution $\mathbf{\b{c}_0}=\mathbf{\b{0}}$, which is incompatible
with  $n_s>0$

Both with $i_0=1$ and $i_0=0$ we obtain that
conditions~(\ref{condo1ns1}) and~(\ref{ns2}) lead to the trivial
solution $\mathbf{\b{c}_0}=\mathbf{\b{0}}$, i.e. $n_s=0$.

A fortiori conditions~(\ref{condo1ns1}) and~(\ref{ns1}) will admit
as only solution the trivial one. This implies
indistinguishability also in the case $n_s=1$.

%\end{multicols}


\begin{thebibliography}{99}
\bibitem{commut}        To demonstrate these commutation relations
write $\left[\tilde{P}_{aux},\tilde{P}_{\psi_i}^\dag\right]=0$ as
a power series of normally ordered $d^\dag$, $d$ operators, and
impose that each term of this power series must vanish. The
coefficients with power $n_a$, $n_a-1$, $n_s$, and $n_s-1$ include
the commutators we want to calculate.
\end{thebibliography}
\end{document}